\documentclass[12pt]{iopart}

\usepackage{iopams}  
\usepackage{graphicx}
\usepackage{multicol}
\usepackage{epstopdf}
\begin{document}

\title[]{Investigation of nonlinear effects in Josephson parametric oscillators used in circuit QED}

\author{Philip Krantz$^{1}$, Yarema Reshitnyk$^{2}$, Waltraut Wustmann$^{1}$, Jonas Bylander$^{1,3}$, Simon Gustavsson$^{3}$, William D. Oliver$^{3,4}$, Timothy Duty$^{5}$, Vitaly Shumeiko$^{1}$, and Per Delsing$^{1}$}

\address{$^{1}$
Microtechnology and Nanoscience, Chalmers University of Technology, Kemivagen 9, SE-41258, Gothenburg, Sweden\\$^{2}$ARC Centre of Excellence for Engineered Quantum Systems, University of Queensland, St Lucia, Queensland 4072, Australia\\$^{3}$Research Laboratory of Electronics, Massachusetts Institute of Technology, Cambridge, Massachusetts 02139, USA\\$^{4}$MIT Lincoln Laboratory, 244 Wood Street, Lexington, Massachusetts 02420, USA\\$^{5}$ARC Centre of Excellence for Engineered Quantum Systems, University of New South Wales, Sydney, New South Wales 2052, Australia}
\ead{philip.krantz@chalmers.se}

\begin{abstract}
We experimentally study the behavior of a parametrically pumped nonlinear oscillator, which is based on a superconducting $\lambda /4$ resonator, and is terminated by a flux-tunable SQUID. We extract parameters for two devices. In particular, we study the effect of the nonlinearities in the system and compare to theory. The Duffing nonlinearity, $\alpha$, is determined from the probe-power dependent frequency shift of the oscillator, and the nonlinearity, $\beta$, related to the parametric flux pumping, is determined from the pump amplitude for the onset of parametric oscillations. Both nonlinearities depend on the parameters of the device and can be tuned \textit{in-situ} by the applied dc flux. We also suggest how to cancel the effect of $\beta$ by adding a small dc flux and a pump tone at twice the pump frequency.
\end{abstract}

\maketitle

\section{INTRODUCTION}

Over the past decade, the interest for parametric systems based on Josephson junctions has revived substantially since the pioneering discoveries\cite{Wahlsten}\cite{Yurke}\cite{Yurke2}\cite{Landau}, due to their implementation in various amplification schemes used to detect weak microwave photons in quantum devices\cite{DykmanPRE1998}\cite{DykmanBook}\cite{Siddiqi2004}\cite{Vijay}\cite{WaltrautPRB}. One appeal of these circuits is associated with the presence of multistable regimes, naturally occuring in nonlinear systems\cite{Strogatz}. Sharp transitions separate these regimes (in phase space), making the devices very useful as sensitive probes of quantum dynamics. When engineering quantum systems in circuit quantum electrodynamics (cQED) architectures, the power of the microwave signal to be measured often reaches the single-photon regime, and consequently the limiting factor of experiments is often the ability to detect and amplify these weak signals with sufficient signal-to-noise ratio.\\
\indent A requirement for the implementation of quantum information processing is to read out the states of quantum bits (qubits) with high fidelity on short time scales compared with the qubit coherence times. In order to coherently manipulate a superconducting qubit as well as to protect it from noise, it is placed in an engineered electromagnetic environment often realized by a superconducting resonator. The combined qubit$\--$resonator system can then be described by the Jaynes-Cummings Hamiltonian\cite{JaynesCummings}. When the qubit transition frequency $\omega_a$ is far detuned from the resonator's angular frequency $\omega_r$, compared to the qubit$\--$resonator coupling rate $g$, $\Delta = \left| \omega_a - \omega_r\right| \gg g$, the resonator picks up a dispersive frequency shift $\omega_r \pm g^2 / \Delta$, with a sign depending on the qubit state. This provides a way to non-destructively probe the qubit dynamics through the resonator response and has been extensively used as a qubit readout method. However, the measurement fidelity is often limited by the weak response signal relative to the noise added from the cryogenic high-electron mobility transistor (HEMT) amplifier.\\
\indent The need to overcome this measurement obstacle has boosted the interest for parametric amplifiers\cite{Yamamoto}\cite{CastellanosBeltran}, which offer a large signal gain and a possibility to continuously probe the system without adding a large amount of noise. Building a parametric amplifier requires a nonlinearity, which in a resonant circuit has the desirable consequence of introducing instabilities and bifurcation points. These add degrees of freedom and complexity to the resonator dynamics, which can be used to implement more efficient readout schemes. The natural and well known candidate as a nonlinear element in superconducting circuits is the Josephson junction, due to its low dissipation and nonlinear inductance. An example of a Josephson-based device utilizing this nonlinearity is the Josephson bifurcation amplifier (JBA) \cite{Siddiqi2004}\cite{Vijay}. It consists of a $\lambda/2$ resonator with a Duffing nonlinearity, realized by placing a Josephson junction in its current node, and has been used to perform single-shot read out of a transmon qubit\cite{Mallet}. In this readout scheme, the two qubit states are brought into correspondence with two oscillation states of the hysteretic bistable system. By probing the resonator close to its bifurcation threshold, the dispersive shift from the qubit state is used to push the resonator into its bistable state where a sharp jump in amplitude is observed for one of the qubit states but not the other. This has enhanced the readout contrast sufficiently to obtain a fidelity of 94$\%$ using a single-shot sample-and-hold pulse sequence of the resonator probe.\\
\indent In this work, we investigate the experimental manifestation of two types of nonlinearities occuring in a nonlinear resonator with a parametrically flux-modulated boundary condition. In addition to the Duffing nonlinearity present in the JBA, the magnetic-flux modulation of the Josephson inductance adds an additional degree of freedom, and a nonlinearity to the system dynamics. The flux modulation enters into the Josephson energy term of the resonator's boundary condition in the form of a mixing product with the field inside the resonator\cite{WaltrautPRB}, $2E_J \left|\cos (\pi \Phi(t)/\Phi_0)\right|\sin (\phi(t))$, where $E_J$ is the Josephson energy, $\Phi(t)$ and $\Phi_0$ are the magnetic flux and flux quantum, respectively, and $\phi(t)$ denotes the phase across the Josephson junctions directly related to the field in the resonator. Considering the Taylor expansions of the flux- and phase contributions to the mixing product\cite{KyleArXiV}, the number of terms entering into the dynamics is set by the microwave pump strength and the number of photons in the resonator.\\
\indent Our measured devices consist of a distributed $\lambda/4$ coplanar waveguide resonator of length $l$, with a flux-tunable inductance realized by terminating one end to ground via two parallel Josephson junctions forming a dc-superconducting quantum interference device, (dc-SQUID)\cite{Yamamoto}\cite{Sandberg}\cite{Sandberg2}\cite{PalaciosLaloy}, see Fig.\@ \ref{Fig1}(a). By threading the SQUID loop with magnetic flux, the electrical length of the resonator is tuned through the changing Josephson inductance

\begin{equation}
L_s = \frac{\Phi_0}{2\pi I_c \left| \cos (\pi \Phi_{\tiny{\mbox{dc}}}/\Phi_0) \right|},
\label{eqSQUIDinductance}
\end{equation}

\noindent where $I_c$ is the critical current of the SQUID. To operate the device parametrically, we modulate the flux around a static dc-bias point, $\Phi_{\tiny{\mbox{dc}}}$, by coupling the SQUID to an on-chip microwave pump line\cite{Sandberg2}, yielding a total flux $\Phi(t) = \Phi_{\tiny{\mbox{dc}}} + \Phi_{1}\cos (\omega_{p}t)$. If this flux pumping is done at around twice the fundamental resonator frequency, $\omega_p \approx 2\omega_r$, parametric oscillations build up the field exponentially in time inside the resonator, above the parametric threshold\cite{WaltrautPRB}\cite{ChrisPRL}. The amplitude of the field in the resonator is eventually limited by the Duffing nonlinearity. The parametric pumping of the boundary condition is the same as that in the dynamical Casimir effect experiment\cite{WilsonDCE}. However, when the field is confined inside a resonator, certain conditions are imposed on the pump-resonator detuning $\delta$ and the effective strength of the pump $\epsilon$ to observe parametric effects. Since oscillations occur only within a limited region in the [$\delta$,$\epsilon$]-plane, it is possible to use such a parametric oscillator as a new member of the family of dispersive read-out techniques for superconducting qubits. In this case, the parametric resonator would work as a threshold detector, in which the two qubit states would be encoded into one oscillating- and one quiet state. This technique would relate to the bifurcation amplifier\cite{Vijay} in the sense of utilizing the cavity pull to push the system into a bistable oscillating state. However, in contrast to the JBA, the resonator can be left empty for one of the two states where the parametric pumping does not build up an oscillating field in the resonator. Depending on the choice of operation point, the choice of "quiet state" can be tailored to reduce back action on the qubit\cite{Picot}, e.g. by encoding the ground state of the qubit into the oscillating state of the resonator.\\
\indent The motivation for this work is to facilitate future designs of pumped nonlinear systems by developing an understanding of these two leading nonlinearities. In particular, when the device is operated as a parametric amplifier, a large bandwidth is preferable. This has the unwanted consequence that the system needs to be parametrically pumped at higher pump strength, introducing higher-order terms in the pump expansion, which we need to account for when operating the system.\\
\indent The structure of the paper is as follows. In sections 2 and 3, we introduce the frequency tunability of the resonator with applied magnetic flux and the theoretical framework for the field inside the resonator, respectively. Next, in section 4, the Duffing nonlinearity is described and extracted, whereas section 5 is devoted to the pump-induced nonlinearity.

\section{TUNABILITY OF THE FUNDAMENTAL FREQUENCY:\\A MEASUREMENT OF THE JOSEPHSON INDUCTANCE}
\begin{figure}[htp] 
\begin{center} 
\includegraphics[width=0.6\columnwidth]{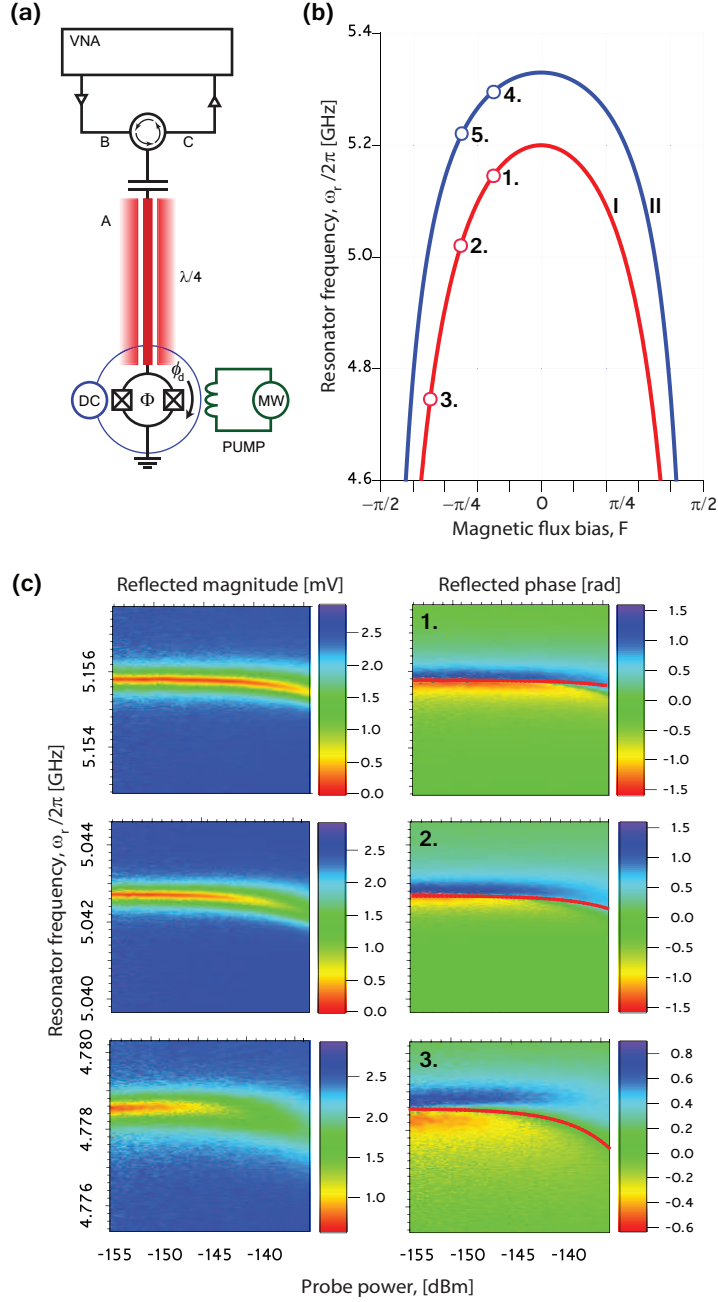} 
\caption{\textbf{(a)} Schematic circuit diagram of the measurement setup using a vector network analyzer (VNA). The quarter-wavelength coplanar waveguide (CPW) resonator (red) is defined by a coupling capacitor to the probe line in one end and shorted to ground via the SQUID in the other. The dc-flux bias $\Phi_{\tiny{\mbox{dc}}}$ is set using a superconducting coil (blue) mounted on the sample box, whereas the microwave-pump, used to modulate the flux around $\Phi_{\tiny{\mbox{dc}}}$, is realized by an on-chip fast tuning line (green). \textit{A} denotes the field inside the resonator. \textit{B} and \textit{C} denote the incoming (probe) and reflected field waves, respectively. \textbf{(b)} Extracted resonant frequencies of the two devices in Table \ref{tab1}, fitted to Eq. (\ref{eqCavityFrequency}), with different inductive participation ratios, $\gamma_0$, yielding slightly different frequency-flux curvatures. \textbf{(c)} Reflected magnitude and phase responses for the three dc flux-bias points for sample I: $F_{1} = -0.15\pi$, $F_{2} = -0.25\pi$, and $F_{3} = -0.35\pi$. The Duffing term gives rise to a nonlinear shift of the resonant frequency as the probe power on the chip is increased. The shift gets more pronounced and affects the resonator at lower probe powers when $F \rightarrow \pm \pi/2$, as indicated by the solid red lines in the reflected phase panels, showing a fit to Eq. (\ref{eqFinalShift}) for parameters presented in Table \ref{tab2}.} 
\label{Fig1} 
\end{center} 
\end{figure}

The first step in characterizing our system's dynamics is to find its fundamental frequency's dependence on the applied dc-flux bias, $F = \pi \Phi_{\tiny{\mbox{dc}}}/\Phi_0$. We measured the devices using a vector network analyzer (VNA) connected to a microwave reflectometry setup, depicted in Fig. \ref{Fig1}(a), in a dilution refrigerator with a base temperature of 20 mK. The shape of the frequency tuning curve as a function of applied magnetic flux is governed by the participation ratio of the SQUID's nonlinear Josephson inductance $L_s$ in Eq. (\ref{eqSQUIDinductance}) to the geometrical resonator inductance, $\gamma_0 = L_{s}(F = 0) / Ll$, where $L$ is the inductance per unit length of the resonator and $l$ its length\cite{Yamamoto}\cite{Sandberg}. The frequency is well approximated by

\begin{equation}
\omega_r (F) \approx \frac{\omega_{\lambda/4}}{1 + \gamma_0/\left|\cos (F)\right|},
\label{eqCavityFrequency}
\end{equation}

\noindent where $\omega_{\lambda/4} = \left. \omega_r\right|_{\gamma_0 = 0}$ denotes the bare resonant frequency, in the absence of the Josephson contribution to its total inductance. Since the nonlinearity originates from the SQUID inductance\cite{WaltrautPRB}, the frequency-flux curvature governs much of the rich nonlinear dynamic properties of the system; $\gamma_0$ and $\omega_r(0)$ should therefore be the main design aspects to consider. To investigate where the nonlinearities enter into the system response, we measured two samples with parameters listed in Table \ref{tab1}. We extracted resonant frequencies and are plotted as a function of magnetic flux in Fig. \ref{Fig1}(b).

\begin{table}[ht]
\caption{Extracted resonator parameters for the two measured samples. $\omega_{\lambda/4}$ and $\omega_{r}(0)$ are the bare- and zero-flux resonant frequencies, respectively. $\gamma_0$ denotes the inductive participation ratio and $I_c$ is the critical current of the SQUID.}
\centering
\begin{tabular}{c c c c c}
\hline
\hline
Sample & $\omega_{\lambda/4}/2\pi$ [GHz] & $\omega_{r}(0)/2\pi$ [GHz] & $\gamma_0$ & I$_c$ [$\mu$A] \\
[0.2ex]
\hline
I & 5.645 & 5.200 & 0.0898 & 2.18\\
II & 5.626 & 5.344 & 0.0563 & 3.48\\
\hline
\end{tabular}
\label{tab1}
\end{table}

\newpage
\section{DIFFERENTIAL EQUATION FOR THE INTRACAVITY FIELD}

The intracavity dynamics of the underdamped, parametrically driven nonlinear oscillator can be mapped onto the Duffing oscillator, studied in detail by M. Dykman \textit{et al.}\cite{DykmanPRE1998}. To investigate the resonator response upon parametric pumping of the flux at frequency $\omega_p$, we adopt the formalism developed by Wustmann and Shumeiko\cite{WaltrautPRB}. Close to resonance, $\delta \equiv \omega_p / 2 - \omega_r \ll \omega_r$, the field amplitude inside the resonator, $A$, can be treated as a slow variable compared to all other timescales in the system, yielding a simplified Langevin equation 

\begin{equation}
i\dot{A} + \delta A + \epsilon A^{*} + \alpha \left| A\right|^2 A + i \Gamma A = \sqrt{2 \Gamma_0} B(t).
\label{eqLangevin}
\end{equation}

\noindent $\left|A\right|^2$ gives the number of photons in the resonator, whereas $B(t)$ is the probe field amplitude such that $\left|B\right|^2$ has units of photons per second. $\Gamma = \Gamma_0 + \Gamma_R$ is the total damping rate of the system, being the sum of the external, $\Gamma_0$, and internal, $\Gamma_R$, damping rates. $\epsilon$ and $\alpha$ denote the effective pump strength and Duffing parameter, respectively. The full $F$-dependence of these coefficients can be express in terms of resonator parameters as\cite{WaltrautPRB}

\begin{equation}
\epsilon \approx \frac{\delta f \omega_{\lambda/4} \gamma_0}{2} \frac{\sin(F)}{\cos^2(F)}
\label{eqEpsilon}
\end{equation}

\begin{equation}
\alpha  \approx \frac{\pi^2 \omega_{\lambda/4} Z_0}{R_K} \left(\frac{\gamma_{0}}{\cos(F)}\right)^3 =  \alpha_0 \left(\frac{\gamma_{0}}{\cos(F)}\right)^{3},
\label{eqAlpha}
\end{equation}

\noindent where $\delta f = \pi \Phi_{1} / \Phi_{0}$ is the ac-flux amplitude, $Z_0 = 50$ $\Omega$ is the resonator's characteristic impedance, $R_K = h/e^2$ is the quantum resistance, and $\alpha_0 = \pi^2 \omega_{\lambda/4} Z_0 / R_K$. In the following two sections we will investigate the Duffing nonlinearity as well as the next order pump-induced nonlinearity.

\section{DUFFING NONLINEARITY} 
\label{sec:Duffing}
The first nonlinearity that we investigate is the effective Duffing parameter, denoted by $\alpha$ in Eq. ($\ref{eqLangevin}$), which we study by probing the resonator with an incoming field $B$ at frequency $\omega_B$, but without parametric pumping, $\epsilon = 0$. This term is also known as the effective Kerr nonlinearity\cite{Mallet}\cite{BertetDykman} and is associated with the cubic term of the intracavity field, related to the current flowing through the SQUID junctions. The magnetic flux dependence of the critical current of the SQUID is therefore projected onto the parameter $\alpha$. The Duffing term gives rise to a nonlinear frequency shift, yielding a resonator lineshape deviating from the linear Lorentzian magnitude response, which can be obtained for weak enough probe power. \\
\indent Experimentally, this nonlinearity can be extracted by measuring the frequency shift of the resonator by probing it with incrementally increasing powers, but far below the parametric instability threshold at which the system bifurcates\cite{Manucharyan}\cite{BertetDykman}, see Fig. \ref{Fig1}(c). In order to extract the parameter $\alpha$ with high precision, a careful calibration of the intracavity field $A$ is necessary. One way to effectively calibrate the field amplitudes and setup attenuation is to use the AC-Stark shift of a sensitive field probe, using a superconducting qubit\cite{BertetDykman}. Here, we will instead estimate the field by determining the resonator damping rates and from that obtain a quantitative understanding for how $\alpha$ depends on the flux bias. The intracavity field can be expressed in terms of the probe field amplitude and the resonator damping rates as

\begin{equation}
A = \frac{\sqrt{2 \Gamma_0}}{\zeta + i \Gamma}B,
\label{eqAvsB}
\end{equation}

\noindent where $\zeta = \delta \omega + \alpha |A|^2$ is the effective resonator$\--$probe detuning, and $\delta \omega = \omega_B - \omega_r$ denotes the detuning between the probe signal and the frequency of the fundamental resonator mode. The reflected power can then be written in terms of this effective detuning

\begin{equation}
\frac{|C|^2}{|B|^2} = 1 - \frac{4 \Gamma_0 \Gamma_R}{\zeta^2 + \Gamma^2},
\label{eqReflectionCoeff}
\end{equation}

\noindent using the amplitude relation $C = B - i\sqrt{2 \Gamma_0}A$. The reflected signal in Eq. (\ref{eqReflectionCoeff}) assumes its minimum at $\zeta = 0$. Thus, we see that the resonance undergoes a nonlinear frequency shift from $\left.\delta \omega \right|_{A=0} = 0$ to $\left.\delta \omega \right|_{A \neq 0} = -\alpha |A|^2$, where $\left| A \right|^2$ is the number of photons in the resonator and $\alpha$ represents the frequency shift per photon. By now substituting this into Eq. (\ref{eqAvsB}), the nonlinear shift can be expressed in terms of the probe power and the resonator damping rates,

\begin{equation}
\delta \omega = -\frac{2\alpha \Gamma_0}{\Gamma^2}|B|^2.
\label{eqFinalShift}
\end{equation}

\noindent We see that to extract $\alpha$, we also need to determine the two damping rates for minimum probe power, at the given flux bias point. Eq. (\ref{eqFinalShift}) shows us that the resonator undergoes a nonlinear frequency shift with increased probe power. However, perhaps more interesting is that the choice of dc-flux bias point, $F$, allows us to tune the Duffing parameter $\alpha$ in-situ within a range from $\alpha_0 \gamma_{0}^3$ and upwards.\\
\indent We extract the Duffing parameter $\alpha$ using Eq. (\ref{eqFinalShift}) at three bias points of sample I, plotted along with the measured reflected phase response in Fig. \ref{Fig1}(c) and listed in Table \ref{tab2}.

\begin{table}[ht]
\caption{Extracted parameters from sample I. $\omega_r$ is the resonator frequency at the three different flux-bias points. $\Gamma_0 = \omega_r / Q_{\tiny{\mbox{ext}}}$ and $\Gamma_R = \omega_r / Q_{\tiny{\mbox{int}}}$ are the external and internal damping rates, respectively, related to their corresponding quality factors $Q_{\tiny{\mbox{ext}}}$ and $Q_{\tiny{\mbox{int}}}$, extracted at low probe power. $\alpha$ represents the Duffing shift per photon.}
\centering
\begin{tabular}{c c c c c c}
\hline
\hline
Flux bias & $\omega_{r} / 2\pi$ & $\Gamma_0 / 2 \pi$ & $\Gamma_R / 2 \pi$ & $\alpha / 2\pi$ & $\alpha/\alpha_0$\\
 $$ & [GHz] & [kHz] & [kHz] & [kHz/photon] & [$\times10^{-3}$]\\
[0.2ex]
\hline
F$_1 = -0.15\pi$ & 5.1558 & 429 & 354 & 108 & 0.996 \\
F$_2 = -0.25\pi$ & 5.0427 & 344 & 310 & 215 & 1.99 \\
F$_3 = -0.35\pi$ & 4.7785 & 482 & 299 & 813 & 7.53 \\
\hline
\end{tabular}
\label{tab2}
\end{table}

\section{PUMP-INDUCED NONLINEARITY}
\label{sec:parametric}
\begin{figure}[htp] 
\begin{center} 
\includegraphics[width = 1\columnwidth]{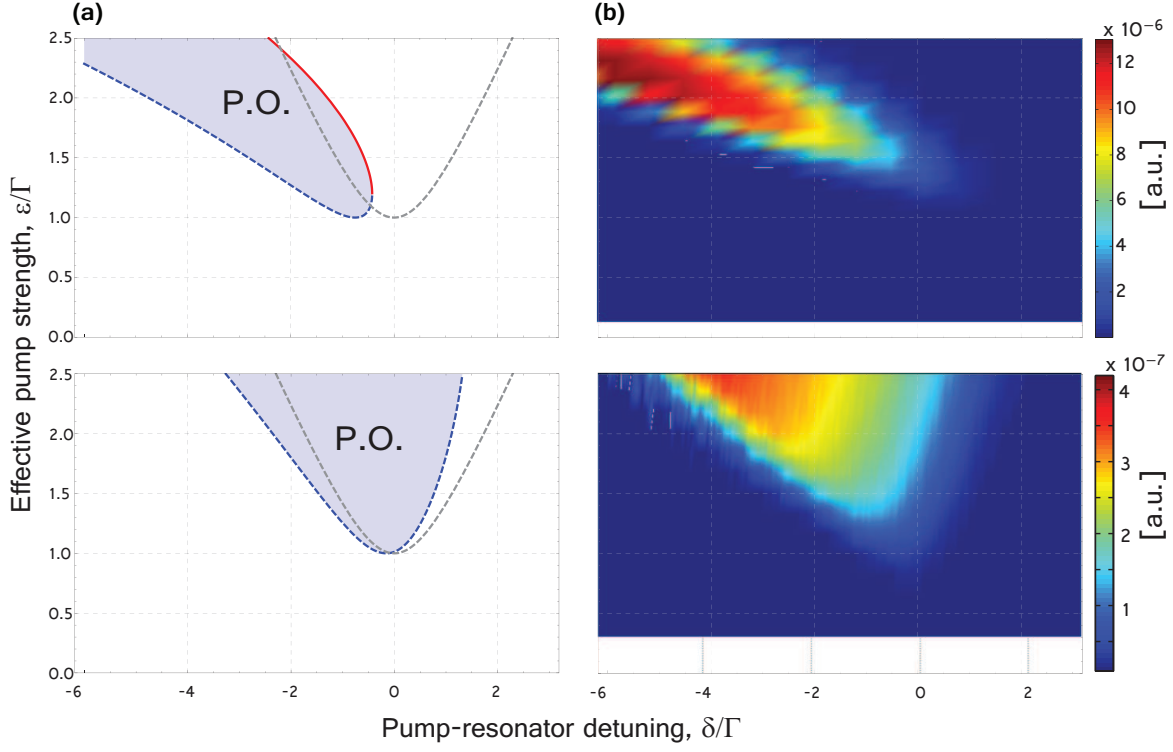}
\caption{\textbf{(a)}Theoretical parametric-oscillation region (P.O.) in the ($\delta \-- \epsilon$)-plane. The dashed blue and solid red lines are the two solutions to Eq. (\ref{eqParaReg}), whereas the dashed gray line indicates the symmetric region in Eq. (\ref{eqParaReg1}), in the absence of a pump-induced frequency shift $\beta$. The two filled theoretical regions are plotted for $\beta_0 = 0.22$. \textbf{(b)} Measured amplitude response around half of the pump frequency at bias points $F_4 = -0.15\pi$ (top) and $F_{5} = -0.25\pi$ (bottom) for sample II, see Fig. \ref{Fig1}b. The faces of the data are interpolated to guide the eye.} 
\label{Fig2} 
\end{center} 
\end{figure}

The next nonlinear effect enters the dynamics when the parametric pumping gets sufficiently strong for higher order terms of the mixing product expansion to affect the resonator. To investigate this nonlinearity, we minimize the Duffing nonlinearity by turning off the probe signal. We then parametrically pump the flux around a bias point a bit higher up on the flux curve where the Duffing influence is weaker, compare Table \ref{tab2}. The first higher-order term is proportional to the square of the pump strength and has the effect of shifting the resonator down in frequency as a consequence of rectification in the flux$\--$frequency transfer function. We reveal this effect by detecting the region of parametric instability in the parameter-plane spanned by the pump$\--$resonator detuning $\delta$ and the effective pump strength $\epsilon$, see Fig. \ref{Fig2}. The energy of the field inside the resonator originates from the pump, and starts to build up exponentially in time when $\epsilon$ is sufficiently strong to compensate for the total damping rate of the resonator: $\epsilon = \Gamma$. After pumping for some time, the field saturates to a steady state set by the Duffing nonlinearity at the given point in the ($\delta \-- \epsilon$)-plane, which we expect to shift the resonator frequency out from the degenerate parametric pumping condition, $\omega_p \approx 2\omega_r$.\\ 
\indent In this section, we will investigate the pump conditions that need to be fulfilled to observe parametric oscillations. The boundaries represent the bifurcation threshold at which the resonator enters into the parametric bistable regime, where oscillations in one of two metastable states of the system Hamiltonian occur\cite{ChrisPRL}. The thresholds are obtained analytically by finding the steady-state, zero-field solutions to the intracavity field differential equation (\ref{eqLangevin})\cite{DykmanPRE1998}\cite{WaltrautPRB}. This yields a threshold symmetric in $\delta$, plotted as the gray dashed line in Fig. \ref{Fig2}(a) and defined by the relation

\begin{equation}
\epsilon = \sqrt{\Gamma^2 + \delta^2}.
\label{eqParaReg1}
\end{equation}

However, this symmetric region does not take into account the pump-induced frequency shift, adding to the detuning of the boundary. This is clearly observed in experiments, see Fig. \ref{Fig2}(b). This effect is a result of the higher pump strength needed to drive parametric oscillations when the first derivative of the frequency-flux curve in Fig. \ref{Fig1}(a) is small compared with the more linear response closer to $F = \pm \pi/2$. This introduces a quadratic, higher order pump term in Eq. (\ref{eqParaReg1}), and the resonator becomes red detuned, which can be understood as a rectification from the deviation from pure sinusoidal pumping of its frequency. We characterize this effect using a dimensionless parameter $\beta$\cite{WaltrautPRB}

\begin{equation}
\omega_{r}(\epsilon) - \omega_r(0) = -\frac{\beta \epsilon^2}{\Gamma}.
\label{eqBeta}
\end{equation}

\begin{figure}[htp] 
\begin{center} 
\includegraphics[scale=1]{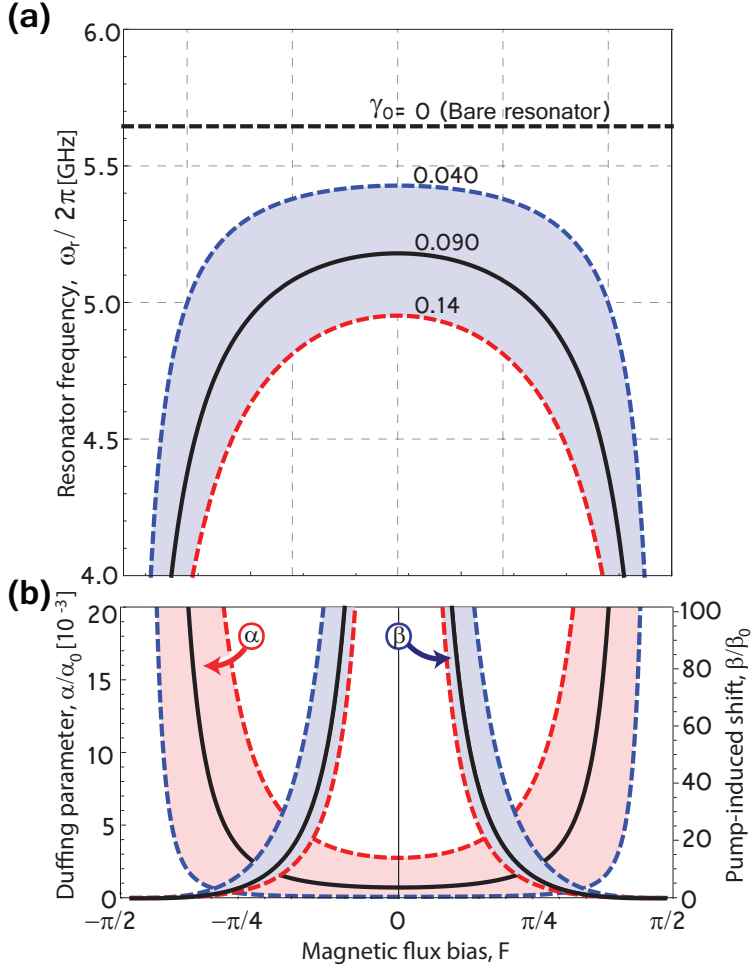} 
\caption{\textbf{(a)} The resonator frequency as a function of magnetic flux from Eq. (\ref{eqCavityFrequency}) for a bare non-tunable resonator ($\gamma_{0} = 0$) (black dashed line) and three values of the inductive participation ratio $\gamma_{0}$ = 0.040, 0.090, and 0.14, in dashed blue, solid black, and dashed red, respectively. \textbf{(b)} Magnetic-flux dependence of the normalized Duffing nonlinearity parameter, $\alpha / \alpha_0$ from Eq. (\ref{eqAlpha}) (left axis) and pump-induced frequency shift parameter, $\beta / \beta_0$ from Eq. (\ref{eqBeta}) (right axis), shown in red and blue regions, respectively. The three different traces correspond to the same values of $\gamma_{0}$ as in (a).} 
\label{Fig3}
\end{center} 
\end{figure}

\noindent When this pump-induced shift is taken into account, we obtain two solutions for the enclosed parametric region, together forming the skewed threshold to the parametric oscillation region in Fig. \ref{Fig2}. The lower and upper parametric instability boundaries follow the relations

\begin{equation}
\frac{\epsilon_{l,u}}{\Gamma} = \frac{1}{\sqrt{2}\beta} \sqrt{1 - 2 \beta \frac{\delta}{\Gamma} \pm \sqrt{1 - 4\beta\left(\beta + \frac{\delta}{\Gamma}\right)}},
\label{eqParaReg}
\end{equation} 

\noindent where the parameter $\beta$ can be approximated in terms of characteristic resonator parameters,

\begin{equation}
\beta \approx \frac{\Gamma}{\omega_{\lambda/4} \gamma_{0}} \frac{\cos^3(F)}{\sin^2(F)} = \frac{\beta_0}{\gamma_{0}}\frac{\cos^3(F)}{\sin^2(F)},
\label{eqBeta} 
\end{equation}

\noindent where $\beta_0 = \Gamma / \omega_{\lambda/4}$. We can develop an intuition for this shift by finding the pump power at which the onset of parametric instability is obtained. This takes place when the pump exactly compensates for the damping and the resonator is empty of photons. The slight shift of the parametric threshold from zero detuning tells us that this is a different effect than the Duffing nonlinearity, which is proportional to the field inside the resonator. Instead, the higher-order pump shift can be understood by considering the curvature of the flux-tuned frequency curve in Fig. \ref{Fig3}(a), plotted for three different values of $\gamma_{0}$. The effective, pump-shifted resonator frequency is lower than the actual static bias point if the second derivative of the curve starts to dominate over the first derivative. Another way to think about this effect is rectification, since the resonator, on average, spends longer time at a lower effective frequency upon parametric pumping due to the steeper curvature on the low-frequency side of the static flux bias point. This is in agreement with the approximation of $\beta$, which diverges as we approach zero flux bias and goes to zero at $F \rightarrow \pm \pi / 2$. This also agrees with the lower pump strength required to drive parametric oscillations for dc-flux bias points at lower frequencies.\\
\indent In Fig. \ref{Fig3}(b), we summarize the flux bias dependence of the Duffing nonlinearity parameter, $\alpha$ in Eq. (\ref{eqAlpha}), described in Sec. \ref{sec:Duffing} and the pump-induced nonlinearity parameter, $\beta$ in Eq. (\ref{eqBeta}). In fact, it is possible to cancel both the rectification and the skewed threshold by adding a dc-component and a second pump tone with a frequency of $2\omega_p$. This is shown in the Appendix A.

\section{CONCLUSIONS}
In conclusion, we have shown how the two different nonlinear effects manifest themselves in superconducting parametric resonators and presented methods to quantify both of them. First, we extracted the Duffing nonlinearity associated with the current flowing through the SQUID by fitting the nonlinear frequency shift as a function of the probe power, using extracted damping rates of the resonator. Second, we studied the parametric response in the absence of a probe signal for two different magnetic-flux points. We conclude that the Duffing nonlinearity dominates at $F\rightarrow \pm \pi/2$, whereas the pump-induced nonlinearity dominates as $F\rightarrow 0$ as a consequence of the fact that the system there needs to be pumped more strongly in order to drive the parametric oscillations, introducing higher order pump terms into the system response. Finally, we see that the interplay between these two nonlinear effects is governed by the inductive Josephson participation ratio, $\gamma_0$. For a small Josephson contribution, the system is more robust against the Duffing shift, but more susceptible to pump-induced nonlinearity and vice versa. With this interplay in mind, more advanced circuits with tailored nonlinear dynamics can be realized. 

\appendix
\section{OPTIMIZATION OF PARAMETRIC PUMPING}
\label{sec:appendixA}

The frequency of a tunable resonator is not linear with respect to the applied magnetic flux. Thus, when supplying a sinusoidal flux, the nonlinear element will experience additional frequency pumping on top of the harmonic base tone. In this appendix, we will demonstrate how this pump-induced deviation from a sinusoidal pump tone effectively can be canceled out by adjusting the pump accordingly. We start out by approximating the frequency tuning using polynomial Taylor expansion around the static dc-flux bias, $F_{\tiny{\mbox{dc}}} = \pi \Phi_{\tiny{\mbox{dc}}} / \Phi_0$

\begin{equation}
\omega(t) \approx \omega\left(F_{\tiny{\mbox{dc}}}\right) + \left.\frac{\partial \omega}{\partial F}\right|_{F = F_{\tiny{\mbox{dc}}}}(F - F_{\tiny{\mbox{dc}}}) + \frac{1}{2}\left.\frac{\partial^2 \omega}{\partial F^2}\right|_{F = F_{\tiny{\mbox{dc}}}}\left(F - F_{\tiny{\mbox{dc}}}\right)^2 + ...
\label{eqApp1}
\end{equation} 

\noindent Using a pure harmonic pump tone $F = F_{\tiny{\mbox{dc}}} + \delta f_1 \cos(\omega_p t)$, yields a frequency-flux relation on the form

\begin{equation}
\omega(t) \approx \omega(F_{\tiny{\mbox{dc}}}) + \omega'(F_{\tiny{\mbox{dc}}})\delta f_1\cos(\omega_p t) +  \frac{\delta f_{1}^2}{4} \omega''(F_{\tiny{\mbox{dc}}})\left(1 + \cos(2\omega_p t)\right)
\label{eqApp2}
\end{equation}

\noindent We see that apart from the fundamental frequency, a dc-rectification contribution, \textit{i.e.} a pump-induced frequency shift proportional to the square of the pump flux amplitude, $\delta f_1$, is added. We also get a second rf tone at twice the pump frequency. \\
\indent Next, we will use this knowledge to adjust the parametric pump signal in such a way that these two higher order pump effects are canceled out. Consider a pump signal ansatz on the following form

\begin{equation}
F(t) = F'_{\tiny{\mbox{dc}}} + \delta f_{1}\cos(\omega_p t) + \delta f_{2}\cos(2 \omega_p t)
\label{eqApp3}
\end{equation}

\noindent where $F'_{\tiny{\mbox{dc}}} = F_{\tiny{\mbox{dc}}} + F_{\tiny{\mbox{rec}}}$. Next, we insert the flux ansatz (\ref{eqApp3}) into the frequency relation in (\ref{eqApp1})

\[
\omega(t) \approx \omega(F'_{\tiny{\mbox{dc}}}) + \frac{\delta f_{1}^2 + \delta f_{2}^2}{4}\omega''(F'_{\tiny{\mbox{dc}}}) + 
\]
\begin{equation}
+ \delta f_{1}\left(\omega'(F'_{\tiny{\mbox{dc}}}) + \frac{\delta f_{2}}{2} \omega''(F'_{\tiny{\mbox{dc}}}) \right) \cos(\omega_p t) +
\label{eqApp4}
\end{equation}
\[
+ \left( \delta f_{2} \omega'(F'_{\tiny{\mbox{dc}}}) + \frac{\delta f_{1}^2}{4} \omega''(F'_{\tiny{\mbox{dc}}})\right)\cos(2\omega_p t) + ...
\]

\noindent Using harmonic balance, the second order tone can be canceled if we satisfy the following condition

\begin{equation}
\delta f_{2} =  - \frac{\delta f_{1}^2}{4} \frac{\omega''(F'_{\tiny{\mbox{dc}}})}{\omega'(F'_{\tiny{\mbox{dc}}})}
\label{eqApp6}
\end{equation}

\noindent By inserting the second order cancelation condition in (\ref{eqApp6}) into the actual pump signal, neglecting higher order, as well as, fast rotating terms\cite{Footnote2}, we get

\begin{equation}
\omega(t) = \omega(F'_{\tiny{\mbox{dc}}}) + \frac{\delta f_{1}^2}{4}\omega''(F'_{\tiny{\mbox{dc}}}) + \delta f_{1}\cos(\omega_p t) \omega'(F'_{\tiny{\mbox{dc}}})
\label{eqApp7}
\end{equation}

\noindent Finally, the dc-rectification component can be evaluated from the ansatz in Eq. (\ref{eqApp3})

\begin{equation}
\omega(F'_{\tiny{\mbox{dc}}}) + \frac{\delta f_{1}^2}{4}\omega''(F'_{\tiny{\mbox{dc}}}) = \omega(F_{\tiny{\mbox{dc}}})
\label{eqApp8}
\end{equation}

\noindent Now, we substitute back the relation for the flux $F'_{\tiny{\mbox{dc}}} = F_{\tiny{\mbox{dc}}} + F_{\tiny{\mbox{rec}}}$ into Eq. (\ref{eqApp8})

\begin{equation}
F_{\tiny{\mbox{rec}}} = - \frac{\delta f_{1}^2}{4}\frac{\omega''(F'_{\tiny{\mbox{dc}}})}{\omega'(F'_{\tiny{\mbox{dc}}})}
\label{eqApp10}
\end{equation}

\noindent In conclusion, to cancel out the second order pump-induced frequency shift, the parametric pumping should be done using the following adjusted signal

\begin{equation}
F(t) = F_{\tiny{\mbox{dc}}} + \delta f_{1}\cos(\omega_p t) - \frac{\delta f_{1}^2}{4}\frac{\omega''(F'_{\tiny{\mbox{dc}}})}{\omega'(F'_{\tiny{\mbox{dc}}})} (1 + \cos(2\omega_p t))
\label{eqApp11}
\end{equation}

\begin{figure}[htp] 
\begin{center} 
\includegraphics[width=0.5\columnwidth]{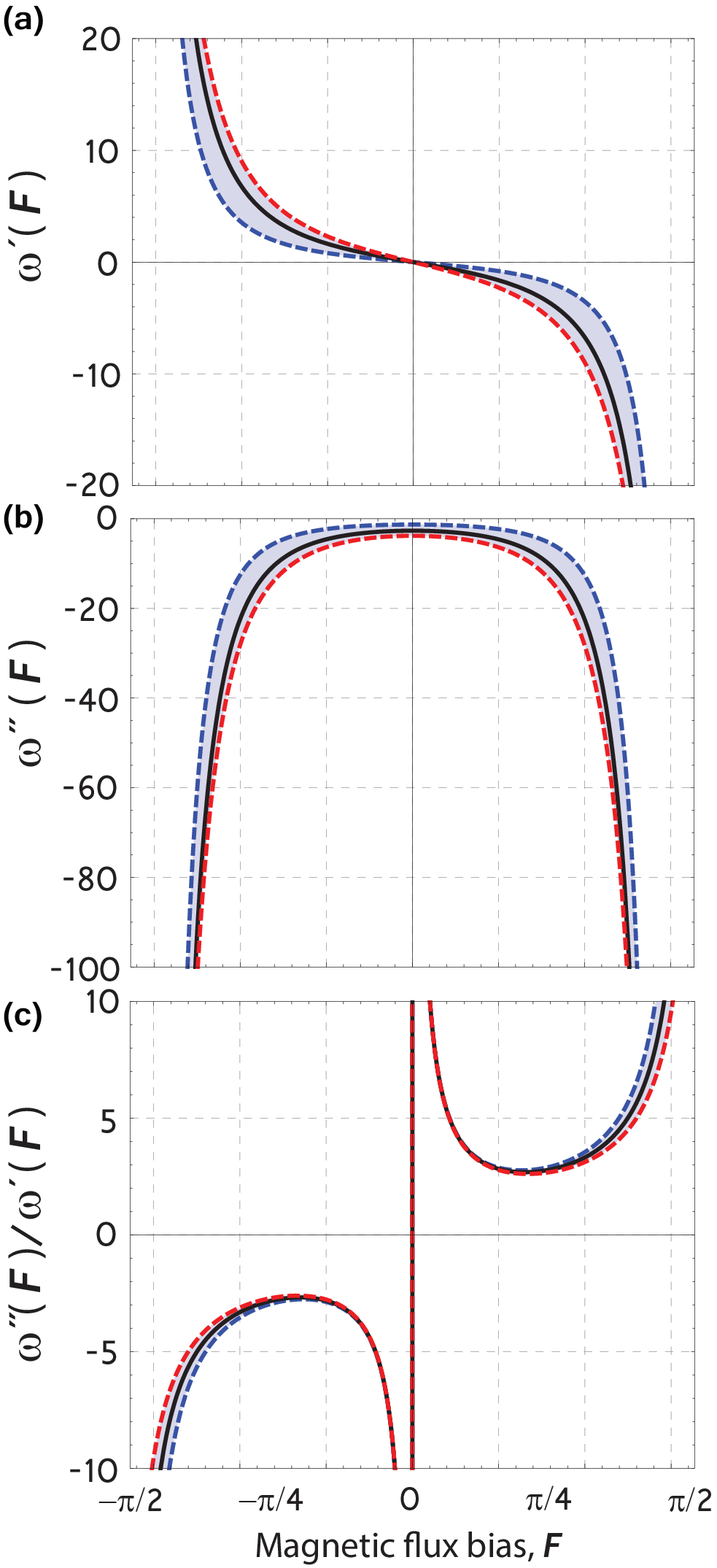} 
\caption{\textbf{(a)} The first derivative of the frequency tuning curve with respect to magnetic flux, $\omega'(F)$, plotted for the same three values of the inductive participation ratio $\gamma_0$ = 0.040, 0.090, and 0.14, plotted with dashed blue, solid black, and dashed red lines, respectively. \textbf{(b)} The second derivative of the frequency with respect to magnetic flux, $\omega''(F)$. \textbf{(c)} The ratio of the second derivative in \textbf{(b)} to the first in \textbf{(a)}, given in equation (\ref{eqApp14})}.
\label{Fig4}
\end{center} 
\end{figure}

Let us now evaluate the cancelation scheme for a tunable resonator with the resonance frequency well approximated by Eq. (\ref{eqCavityFrequency}), where the inductive participation ratio of the system is small ($\gamma_0 \ll 1$). To find the compensation terms, we derive the first and second derivatives of the frequency with respect to magnetic flux, plotted in Fig. \ref{Fig4}, together with the ratio of the second derivative to the first, given by

\begin{equation}
\frac{\omega''(F'_{\tiny{\mbox{dc}}})}{\omega'(F'_{\tiny{\mbox{dc}}})} = \frac{3 + 2\gamma_0 \cos (F'_{\tiny{\mbox{dc}}}) + \cos (2F'_{\tiny{\mbox{dc}}})}{2 \sin (F'_{\tiny{\mbox{dc}}})(\gamma_0 + \cos(F'_{\tiny{\mbox{dc}}}))}
\label{eqApp14}
\end{equation}

\noindent In conclusion, the pumping needed to compensate for the second order pump-induced nonlinearity in a tunable resonator can be written on the following form

\[
F(t) = F_{\tiny{\mbox{dc}}} + \delta f_{1} \cos \left(\omega_p t\right) - 
\]
\begin{equation}
\indent - \frac{\delta f_{1}^2}{4}\left(\frac{3 + 2\gamma_0 \cos (F'_{\tiny{\mbox{dc}}}) + \cos (2F'_{\tiny{\mbox{dc}}})}{\sin (F'_{\tiny{\mbox{dc}}})(\gamma_0 + \cos(F'_{\tiny{\mbox{dc}}}))}\right) \left( 1 + \cos (2\omega_p t)\right)
\label{eqApp15}
\end{equation}

\section*{ACKNOWLEDGEMENTS}
The samples were fabricated in the MC2 clean room facilities at Chalmers University of Technology, Sweden and measured at University of Queensland, Australia. We acknowledge financial support from the European Research Council, ERC No. 247208, the Australian Research Council Centre of Excellence in Engineered Quantum Systems, ARC No. CE110001013, and the Swedish Foundation for International Cooperation in Research and Higher Education, STINT No. IG2009-2023. We gratefully thank C. M. Wilson, L. Tornberg, I.C. Hoi, and M. Simoen for fruitful discussions.


\newpage
\section*{REFERENCES}
\begin{thebibliography}{10}
\bibitem{Wahlsten} Wahlsten S \textit{et al.} 1977 \textit{Appl. Phys. Lett.} \textbf{30} 298
\bibitem{Yurke} Yurke B \textit{et al.} 1988 \textit{Phys. Rev. Lett.}, \textbf{60} 764-767
\bibitem{Yurke2} Yurke B \textit{et al.} 1989 \textit{Phys. Rev. A}, \textbf{39} 2519
\bibitem{Landau} Landau L D and Lifshitz E M 1976 \textit{Mechanics} vol. 1, 3rd edn., (Oxford: Pergamon)
\bibitem{DykmanPRE1998} Dykman M 1998 \textit{Phys. Rev. E}, \textbf{57} 5202
\bibitem{DykmanBook} Dykman M 2012 \textit{Fluctuating Nonlinear Oscillators} ed M Dykman (Oxford: Oxford University Press) pp 165-197
\bibitem{Siddiqi2004} Siddiqi I \textit{et al.} 2004 \textit{Phys. Rev. Lett.} \textbf{93} 207002
\bibitem{Vijay} Vijay R \textit{et al.} 2009 \textit{Rev. Sci. Inst.} \textbf{80} 111101
\bibitem{WaltrautPRB} Wustmann W and Shumeiko V S 2013 \textit{Phys. Rev. B} \textbf{87} 184501
\bibitem{Strogatz} Strogatz S H 1994 \textit{Nonlinear dynamics and chaos} (Reading, MA: Addison-Wesley)
\bibitem{JaynesCummings} Jaynes E T and Cummings F W 1963 \textit{Proc. IEEE} \textbf{51} 89-109
\bibitem{Yamamoto} Yamamoto T \textit{et al.} 2008 \textit{Appl. Phys. Lett.} \textbf{93} 042510
\bibitem{KyleArXiV} Sundquist K \textit{et al.} 2013 \textit{Appl. Phys. Lett.} \textbf{103} 102603
\bibitem{CastellanosBeltran} Castellanos-Beltran M A and Lehnert K W 2007 \textit{App. Phys. Lett.} \textbf{91} 083509
\bibitem{Mallet}Mallet F \textit{et al.} 2009 \textit{Nature Physics} \textbf{5} 791-795
\bibitem{Wallquist2006} Wallquist M, Shumeiko V S and Wendin G 2006 \textit{Phys. Rev. B} \textbf{74} 224506
\bibitem{Sandberg} Sandberg M \textit{et al.} 2009 \textit{Physica Scripta} \textbf{T137} 014018
\bibitem{Sandberg2} Sandberg M \textit{et al.} 2008 \textit{Appl. Phys. Lett.} \textbf{92} 203501
\bibitem{ChrisPRL} Wilson C M \textit{et al.} 2010 \textit{Phys. Rev. Lett.} \textbf{105} 233907
\bibitem{PalaciosLaloy} Palacios-Laloy A \textit{et al.} 2008 \textit{J. Low Temp. Phys.} \textbf{151} 1034-42
\bibitem{Picot} Picot T \textit{et al.} 2008 \textit{Phys. Rev. B}, \textbf{78} 132508
\bibitem{WilsonDCE} Wilson C M \textit{et al.} 2011 \textit{Nature} \textbf{479} 376-379
\bibitem{Manucharyan} Manucharyan V E \textit{et al.} 2007 \textit{Phys. Rev. B}, \textbf{76} 014524
\bibitem{BertetDykman} Bertet P \textit{et al.} 2012 \textit{Fluctuating Nonlinear Oscillators} ed M Dykman (Oxford: Oxford University Press) pp 1-32

\end{thebibliography}
\end{document}